# Clustering in the 1.2 Jy *IRAS* Galaxy Redshift Survey I: The Redshift and Real Space Correlation Functions*

Karl B. Fisher[1], Marc Davis[2], Michael A. Strauss[3], Amos Yahil[4], and John Huchra[5]
[1] *Institute of Astronomy, Madingley Rd., Cambridge, CB3 0HA, England*
[2] *Astronomy and Physics Departments, University of California, Berkeley, CA 94720*
[3] *Institute for Advanced Study, School of Natural Sciences, Princeton, NJ 08540*
[4] *Astronomy Program, State University of New York, Stony Brook, NY 11794-2100*
[5] *Center for Astrophysics, 60 Garden St., Cambridge, MA 02138*

**ABSTRACT**
We present analyses of the two-point correlation function derived from an all-sky redshift survey of 5313 galaxies extracted from the Infrared Astronomical Satellite (*IRAS*) database. The redshift space correlation function $\xi(s)$ is well described by a power law, $\xi(s) = (s/4.53 h^{-1} \mathrm{Mpc})^{-1.28}$, on scales $\lesssim 20\ h^{-1}\mathrm{Mpc}$; on larger scales $\xi(s)$ drops below the extension of this power law. We examine the effect of redshift space distortions on the correlation function and compute the full two dimensional correlation function $\xi(r_p, \pi)$. From this, we derive the real space correlation function, which is well described by $\xi(r) = (r/3.76 h^{-1} \mathrm{Mpc})^{-1.66}$ on scales $\lesssim 20\ h^{-1}\mathrm{Mpc}$.
The derived correlation functions are found to be consistent with previous determinations in the literature, and seem to show more power on large scales than predicted by the standard Cold Dark Matter (CDM) model. Comparison of the derived $\xi(r)$ with the correlation function of optical galaxies implies an optical to *IRAS* bias ratio of $b_O/b_I = 1.38 \pm 0.12$ on a scale of $\sim 8\ h^{-1}\mathrm{Mpc}$. The variances in cubical cells inferred from $\xi(s)$ appear discrepant with the previously reported results of Efstathiou *et al.* (1990).

## 1 INTRODUCTION

Historically, the spatial two-point correlation function, $\xi(r)$, has played an important role in quantifying the clustering of galaxies. Its success stems both from the ease with which it can be computed from existing data, and its direct physical interpretation within the context of gravitational instability theories[†]. It has long been known that $\xi(r)$ is well described by a power law, $\xi(r) = (r/r_0)^{-\gamma}$ on scales $\lesssim 10\ h^{-1}\mathrm{Mpc}$ where the galaxy distribution is characterized by strong nonlinear clustering (e.g., Davis & Peebles 1983). Theoretically, $\xi(r)$ is simply the Fourier conjugate of the power spectrum characterizing the present day galaxy distribution; thus in principle, knowledge of $\xi(r)$ gives vital information about the underlying fluctuations which give rise to the observed galaxy distribution.

Unfortunately, in practice, observations of $\xi(r)$ have yielded only limited information about these underlying fluctuations. Linear perturbation theory predicts that $\xi(r)$ is the Fourier conjugate of the *initial* power spectrum only in the regime where $\xi(r)$ is much less than unity, and we lack a quantitative theory which allows $\xi(r)$ to be predicted in the regime where $\xi(r) \gg 1$ (although see Hamilton *et al.* 1991). However, it has proven extremely difficult to measure $\xi(r)$ from existing redshift surveys on scales large enough for linear theory to apply. This is simply a reflection of statistics; it is difficult to measure correlation

---

amplitudes below the fractional uncertainty in the mean density of the galaxy sample (roughly speaking, with the standard estimator for the correlation function (e.g., Davis and Peebles (1983)) an error in the mean density, $\Delta n$, translates directly into an error in the correlation function, with $\Delta \xi \approx \Delta n/n$; cf., the discussion in Fisher *et al.* 1993*a* although see Hamilton (1993) for a discussion of estimators which are less sensitive to density errors).

Redshift surveys extracted from the *IRAS* database are well suited for the determination of the correlation function on large scales because they cover a large, near full sky, volume with uniform selection criteria, allowing an accurate determination of $n$. In this paper, we investigate clustering via correlation statistics in a sample of 5313 *IRAS* galaxies complete to a flux limit of 1.2 Jy at 60 $\mu$m. This paper is the first of a two part series investigating the clustering in the 1.2 Jy sample using correlation statistics and is the fifth paper based on the analysis of the 1.2 Jy sample. Fisher *et al.* (1992) describe tests for galaxy evolution in the sample. The acceleration of the Local Group is treated in Strauss *et al.* (1992*c*), and Fisher *et al.* (1993*a*) analyze the power spectrum of the 1.2 Jy sample. Finally, Bouchet *et al.* (1993) explore higher order moments of the density field, and their relations with one another. The second paper in this series (Fisher *et al.* 1993*b*) presents an analysis of redshift distortions in galaxy clustering based on two point correlation statistics.

Galaxy candidates were chosen from the *IRAS* Point Source Catalog, Version 2, (1988) using the selection criteria described in Strauss *et al.* (1990) and Fisher (1992). The data for the brighter half of the sample can be found in Strauss *et al.* (1992*b*). At present, thirty objects (0.5% of the sample) remain unobserved. Sky coverage is complete for $|b| > 5°$ with the exception of a small region of the sky which *IRAS* failed to survey and regions limited by confusion; our sample covers 87.6% of the sky. All heliocentric redshifts are converted to the Local Group reference frame using the transformation of Yahil, Tammann, & Sandage (1977). In the remainder of this Paper, the term "redshift" will refer to recession velocities as measured in the rest frame of the Local Group, and no further corrections for peculiar motions will be made to individual peculiar velocities.

This paper is organized as follows. In § 3, we present the *IRAS* correlation function in redshift space $\xi(s)$ using the methods discussed in § 2, and then give a qualitative comparison of the results with various models of structure formation. We examine in §4 redshift space distortions by computing the correlations as a function of both radial ($\pi$) and tangential ($r_p$) separations, $\xi(r_p, \pi)$; we obtain the *real space* correlation function from a projection of $\xi(r_p, \pi)$. In §5 we compute the variance of the galaxy counts inferred from the correlation function, and compare the results with previous determinations from *IRAS* and optically selected samples of galaxies. We conclude in §6.

We have included an appendix that discusses a method for fitting models to $\xi(r)$ which correctly accounts for the covariance between the estimated values at different separations.

## 2 MEASURING THE TWO-POINT CORRELATION FUNCTION

### 2.1 Method

The two point correlation (or equivalently the autocorrelation) function, $\xi(r)$, is defined as the probability in excess of Poisson of finding a galaxy in a volume $\delta V$ a distance $r$, away from a randomly chosen galaxy,

$$\delta P = n \delta V \left[ 1 + \xi(r) \right] \quad , \tag{1}$$

where $n$ is the mean number density of galaxies. In order to calculate $\xi(r)$, one first generates in the computer a sample of points (the random sample) from a uniform distribution, with the same selection criteria as the galaxy sample. A reliable and robust estimator for $\xi(r)$ which is unaffected by the boundaries of the sample is then given by (e.g., Blanchard & Alimi 1988, although see Landy & Szalay 1993),

$$\xi(r) = \frac{N_{DD}(r)}{N_{DR}(r)} \frac{n_R}{n_D} - 1 \quad , \tag{2}$$

where $N_{DD}$ and $N_{DR}$ refer to the number of data-data and data-random pairs, respectively, in a narrow interval of separations centered on $r$. $n_D$ and $n_R$ denote the mean densities of the real and random catalogs, respectively.

The estimator in Equation 2 can be generalized to include an arbitrary weighting function,

$$N_{DD} \text{ or } N_{DR}(\tau) = \sum_{\tau - \Delta\tau/2 < |\mathbf{r}_i - \mathbf{r}_j| < \tau + \Delta\tau/2} w(r_i, \tau)\, w(r_j, \tau) \quad , \tag{3}$$

where the weight $w(r_i, \tau) w(r_j, \tau)$ can depend both on the distance of the objects from the the origin ($r_i, r_j$), and the distance of the two objects from each other ($\tau \equiv |\mathbf{r}_i - \mathbf{r}_j|$). The simplest choice of weights in Equation 3 is to set $w(r_i, \tau) = 1$; this is the choice we adopt for the analysis of volume limited subsamples described in the next section. Saunders *et al.* (1992) have shown that for flux limited samples, the variance in the estimate of $\xi(r)$ on large spatial scales is minimized if $w(r_i, \tau)$ in Equation 3 is

$$w(r_i, \tau) = \frac{1}{1 + 4\pi n_D J_3(\tau) \phi(r_i)} \quad , \tag{4}$$

where $\phi(r)$ is the selection function of the sample with redshifts (such that for a homogeneous distribution of galaxies, the number density of objects in the sample is $n_D \phi(r)$), and $J_3(\tau) \equiv \int_0^\tau dr\, r^2 \xi(r)$. Equation 4 has reasonable properties. In the limit of galaxy pairs at small distances, $r$, we have $4\pi n_D J_3(\tau)\phi(r) \gtrsim 1$, so $w \gtrsim 1/\phi(r)$, giving equal volume weighting to the pairs; this is desirable since these galaxies are well sampled and most pairs are not independent. On the other hand, distant galaxies have $\phi(r) \ll 1$, and structures are not well sampled. Consequently, in this limit, each galaxy is assigned equal weight, i.e, $w \to 1$, in order to avoid excessive weighting of undersampled zones.

Yahil *et al.* (1991) describe our method for finding the selection function $\phi(r)$; we use the same parameterization as they do:

$$\phi(r) = \begin{cases} 1 & \text{if } r < r_m, \\ \left(\frac{r}{r_m}\right)^{-2\alpha} \left(\frac{r_*^2 + r^2}{r_*^2 + r_m^2}\right)^{-\beta} & \text{otherwise,} \end{cases} \tag{5}$$

where $r_m = 635$ km s$^{-1}$; our best solution in redshift space is

$$\alpha = 0.51, \quad \beta = 1.84, \quad r_* = 5440 \text{ km s}^{-1} \quad . \tag{6}$$

Note that the weights $w$ depend on the unknown correlation function via $J_3(\tau)$. In § 2.3, we test the optimal weighting scheme using mock *IRAS* catalogs drawn from $N$-body simulations of a CDM universe; in this case we simply use the $J_3(\tau)$ computed from the linear theory CDM $\xi(r)$. To compute the actual *IRAS* correlation function, we use the real space *IRAS* correlation function determined by Saunders *et al.* (1992), who found

$$\xi(r) = (r/3.79 h^{-1}\text{Mpc})^{-1.57} \text{ for } r \lesssim 30\, h^{-1}\text{Mpc} \quad . \tag{7}$$

On scales larger than $30\, h^{-1}$Mpc, we have set $\xi(r)$ to zero. The resulting $J_3$ is given by

$$J_3(\tau) = \begin{cases} 5.66(\tau/1h^{-1}\text{Mpc})^{1.43}\, (h^{-1}\text{Mpc})^3, & \text{if } \tau < 30\, h^{-1}\text{Mpc}, \\ 724\, (h^{-1}\text{Mpc})^3, & \text{otherwise} \end{cases} \tag{8}$$

Ideally one should compute $\xi(r)$ iteratively, i.e., assume an initial guess for $\xi(r)$, then compute $\xi(r)$, update the weights and repeat until convergence is reached. We do not do this for two reasons: first, the estimator of the correlation function in Equation 2 is unbiased no matter what the weighting function $w$ is; second, the $\xi(r)$ we determine in § 3 is in good agreement with Equation 7.

In our analysis, we use the minimum variance density estimator described in Davis & Huchra (1982) to compute the densities $n_D$ and $n_R$ which appear in Equation 2,

$$\frac{n_R}{n_D} = \frac{\sum_{i=1}^{N_R} w(r_i, \tau = 30\, h^{-1}\text{Mpc})}{\sum_{j=1}^{N_D} w(r_j, \tau = 30\, h^{-1}\text{Mpc})} \quad . \tag{9}$$

The density $n$ which appears in the expression for $w$ again ideally should be calculated iteratively; we simply calculate $n = \frac{1}{V}\sum_i \frac{1}{\phi(r_i)}$, which is close to the minimum variance estimate (Yahil *et al.* 1991). The sums in Equation 9 are over all objects within a sphere of radius $200\, h^{-1}$Mpc.

Figure 1 shows the "effective" window function defined as the weight assigned per unit volume, i.e., $\propto n_D \phi(r) w(r,\tau)$, for the weighting scheme given in Equation 4 and the choice of $J_3$ given in Equation 8. The effective window function gives more weight to distant particles as $\tau$ increases, as is desirable, since large scale correlation estimates require large volumes and are not as affected by dilute sampling. For all values of $\tau$, the window function declines for large values of $r$, minimizing the statistical noise arising from the dilute sampling of galaxies at large distances in a flux limited catalog. The dashed lines in Figure 1 indicate the value of $r$ at which the window function is half its maximum value. This "half power" point is indicative of the effective depth of the weighted sample. The half power point of the window function is limited to $\lesssim 120\, h^{-1}$Mpc, even when the pair separation, $\tau$, is as large as $30\, h^{-1}$Mpc.

In order to avoid objects whose redshifts are dominated by their peculiar velocities, we omit objects in the sample with redshifts $< 500$ km s$^{-1}$. The analysis of the optimally weighted correlation function is computed for all galaxies with redshifts $< 30,000$ km s$^{-1}$. We have used random catalogs with 50,000 artificial galaxies in calculations involving the full flux limited catalog and catalogs with 5000 artificial galaxies for calculations involving volume-limited subsamples of the data. These numbers are sufficiently large to ensure that this is a negligible source of errors in the inferred $\xi(r)$, while small enough to make the calculations computationally feasible.

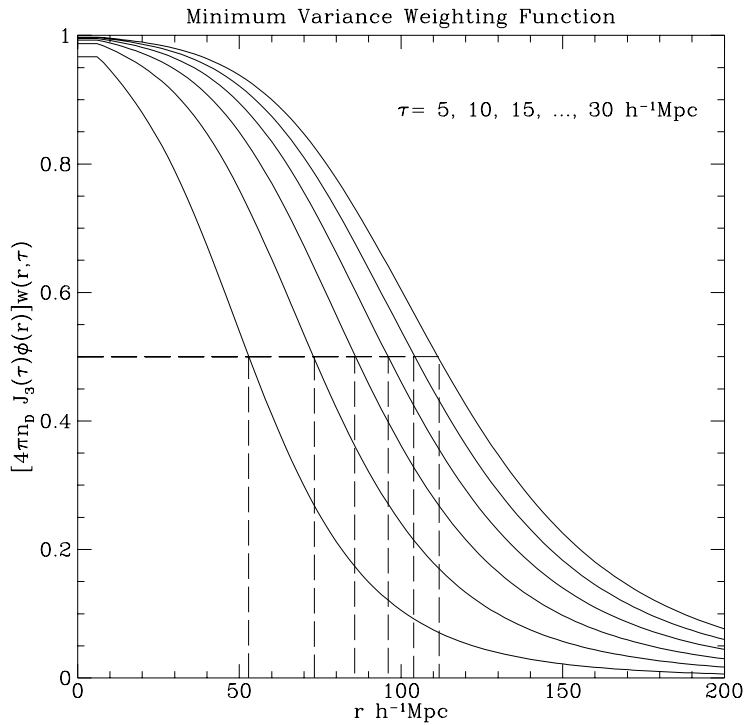

**Figure 1.** The effective window function for $\tau = 5,\ 10,\ \ldots,\ 30\ h^{-1}$ Mpc and the $J_3(\tau)$ given Equation 8. The amplitudes of the curves increase with increasing $\tau$.

## 2.2  Estimation of Statistical Errors

In order to perform fits to the correlation function $\xi$, we need to assess errors on $\xi$ as a function of position; moreover, because the values of $\xi$ at different separations are correlated, we also need the off-diagonal terms in the covariance matrix (Appendix A). The standard method for computing correlation function errors is through the technique of bootstrap resampling (Ling, Frenk, & Barrow 1986). With this technique, one creates a series of $N$ bootstrap data sets of the same size as the original data set by randomly assigning positions to the bootstrap galaxies from the list of positions in the original data (sampling with replacement); consequently some of the galaxy positions in the original data set will occur several times in a given bootstrap data set while others will not be represented. Since the assignment of positions in the bootstrap data sets is performed randomly, the mean of the correlation functions determined from the series of bootstrap data sets will approach the correlation function of the original data set. However, although the bootstrap method gives an unbiased estimator of the underlying distribution of any point process in the limit of large $N$ (Efron 1981; Lupton 1993), there are certain statistics for which the bootstrap will not properly estimate the variances (e.g., Rubin 1981; Press *et al.* 1992). For correlation statistics, there are two problems. First, if the correlation function is computed in a small volume, then it may be very different from the global or *ensemble* mean value of the correlation function predicted by a given theory, simply because the volume of space probed may not be representative of the universe as a whole. This component of the error can be calculated given a model for the underlying power spectrum, either analytically (in which case, one needs to know the three and point correlation functions as well; Mo, Jing, & Börner 1992), or from an $N$-body simulation; see below. In any case, this ensemble error, arising from the finite computational volume, is not reflected in the bootstrap error, and the bootstrap error is an underestimate.

The second problem is due to the sparse sampling by galaxies of the underlying density distribution which we assume the galaxies trace. The problem can best be illustrated by imagining using bootstrap methods to assess the statistical significance of a void seen in the galaxy distribution. No matter how many bootstrap realizations are done, the void will remain empty, thus *underestimating* the error associated with the density estimate (Santiago & Strauss 1992). A similar argument shows that discreteness effects cause bootstraps to *overestimate* the variance in the density field estimate in overdense regions, and as the derived correlation function is heavily weighted by the densest regions, we predict that the bootstrap errors will be an overestimate of the true errors.

This point can be illustrated as follows. Following Itoh *et al.* (1992), we compute ensemble errors for the correlation function from ten mock *IRAS* catalogues drawn from a single realization of an unbiased CDM simulation (Frenk *et al.* 1990), which closely mimic the *IRAS* 1.2 Jy survey in number density, excluded zones, selection function, and clustering properties (Górski *et al.* 1989; Davis *et al.* 1991). For each catalog, we compute the redshift space correlation function using the optimal weighting scheme described in the previous section. In addition, we perform 100 bootstrap resamplings for each of the ten mock samples.

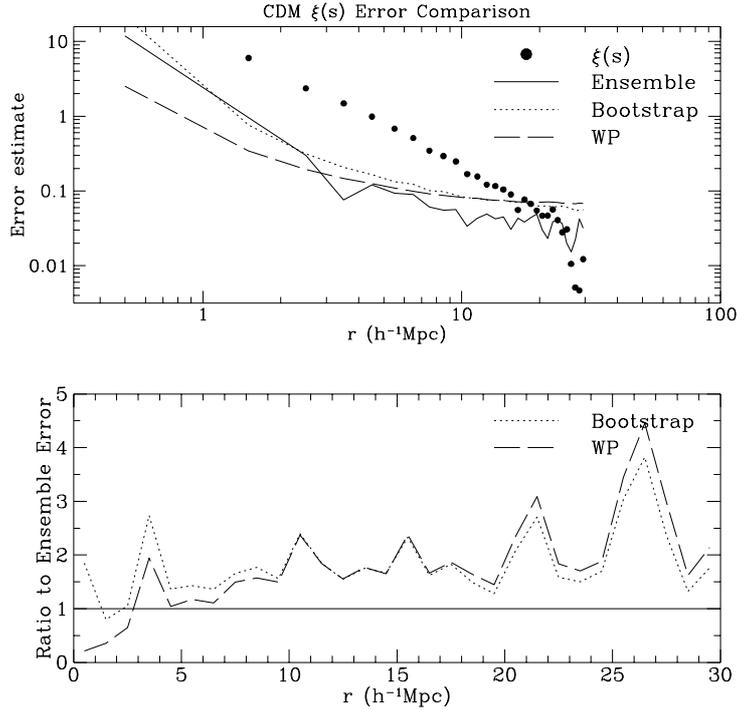

**Figure 2.** Comparison of error estimates from the mock *IRAS* CDM catalogs. The solid line shows the ensemble error computed from the scatter in the different catalogs while the dotted and dashed lines denote the bootstrap and weighted Poisson (WP) errors respectively. The ensemble error curve is much noisier than the other two, as it is based on only ten realizations. The solid points denote the mean CDM redshift space $\xi(s)$. The lower panel shows the ratio of the bootstrap and weighted Poisson errors to the ensemble errors. The impression that the signal-to-noise ratio drops below unity at $\sim 15\ h^{-1}$ Mpc is an artifact of the small bins at large separations.

The standard deviation of the determined correlation function from each mock *IRAS* sample is a measure of the ensemble error; it is given by the solid curve in Figure 2. The bootstrap error averaged over the ten samples is shown in Figure 2 by the dotted curve; it is much smoother than the ensemble errors because it is based on 100 times more realizations. Finally, the "weighted Poisson" error at separation $\tau$ is defined as $\mathrm{WP}(\tau) \equiv (1 + \xi(\tau))(N_{DD}(\tau))^{-1/2}$, where $N_{DD}(\tau)$ is the weighted number of pairs at that separation (cf., Equation 3); this is shown as the dashed curve in the figure. The lower panel shows ratios of the bootstrap and weighted Poisson errors to the ensemble errors. The bootstrap and weighted Poisson errors agree very well on large scales, but both overestimate the ensemble error by roughly a factor of two. The realizations from which the ensemble errors were derived were drawn from a single $N$-body realization of the CDM power spectrum, which will cause the ensemble errors to be somewhat underestimated (Fisher *et al.* 1993a). Ideally one would like to compute the ensemble errors with an ensemble of independent simulations but this is computationally impractical. However, the bulk of the discrepancy between the ensemble and bootstrap errors is due to the bias of bootstrap methods for errors in correlation statistics.

Given this, our ideal approach would be to do a large series of $N$-body simulations of a model with power spectrum well matched by the real data, and use them to define errors and covariances in the correlation functions (cf., Fisher *et al.* 1993a). This is computationally infeasible, and thus we use bootstrap errors and covariances (cf., Appendix A) throughout this paper (except in §2.3 below), keeping in mind that they are probably overestimates of the true error.

### 2.3 Tests of the Method

We now test our technique, and search for possible systematic errors, by determining the correlation function in a sample for which we know the correct correlation function, namely one drawn from an $N$-body simulation. We use the mock *IRAS* simulations drawn from CDM simulations discussed in the previous section. In these catalogs, the full nonlinear real space correlation function is accurately known from pair counts using the full simulation. For this test, the mock catalogs were in real space and therefore not distorted by redshift space effects.

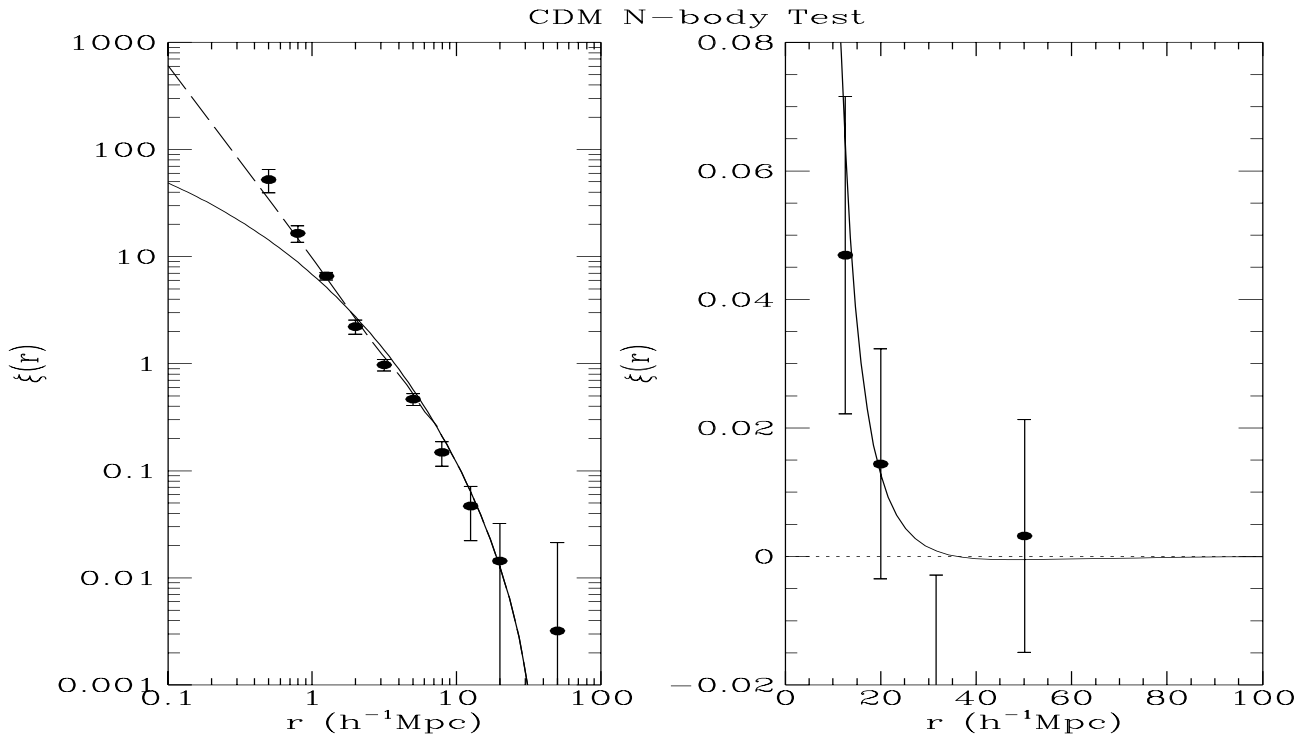

**Figure 3.** Test of method using volume-limited subsamples; samples were volume-limited to 6000 km s$^{-1}$. Dots represent the mean of ten mock *IRAS* catalogs drawn from a CDM *N*-body simulation. Error bars represent the standard deviation of the mean over the ten catalogs. The solid line is the linear theory prediction while the dashed line is the full nonlinear $\xi(r)$.

As our first check, we extracted subsamples volume-limited to 60 $h^{-1}$Mpc from the mock catalogs; the average of the resulting $\xi(r)$ over the ten realizations is shown in Figure 3 (dots) along with the linear theory prediction (solid line) and that calculated from pair counts of the full *N*-body simulation (dashed line). The mean correlation function from the ten observers recovers the true $\xi(r)$ in an unbiased way on all scales. However, the statistical uncertainties in $\xi(r)$ become very large when $\xi(r) \ll 1$. We then proceeded to compute $\xi(r)$ using flux-limited subsamples and the minimum variance weights given in Equation 4; we used the linear theory CDM $\xi(r)$ to compute the $J_3(r)$ needed in the weighting function. The results of this test are shown in Figure 4. We found that the inferred $\xi(r)$ was once again consistent with the true $\xi(r)$ on all scales. Notice in particular that the scatter in the estimates of $\xi(r)$ (the error bars in Figures 3 and 4) is greatly reduced when $\xi(r)$ is computed from the full flux limited catalog rather than from volume limited subsamples; although the intrinsic sample to sample variations are identical in the two cases, the statistical uncertainties in the estimates of $\xi(r)$ are reduced significantly both by the optimal weighting strategy and by the increased number of galaxies in the flux limited catalog.

## 3 THE IRAS REDSHIFT SPACE CORRELATION FUNCTION

### 3.1 Volume Limited Subsamples

The use of volume limited subsamples to compute $\xi(s)$ is very straightforward via Equation 2; the results are easy to interpret because the galaxies used are selected from a relatively narrow range of luminosities. As we saw in the previous section, however, the volume limiting procedure discards a large portion of the available data and hence information on the correlation function. The reduced number of galaxies in the volume limited samples increases the uncertainty in the background density estimate, making it difficult to determine $\xi(r)$ on scales $\gtrsim 20$ $h^{-1}$Mpc.

The subsample volume-limited to a depth $R$ contains all those objects in the full flux limited sample with redshift $< R$ and whose luminosity is greater than that of an object at the flux limit at redshift $R$. We compute the luminosity of each galaxy using the full relativistic and color correction machinery described in Fisher *et al.* (1992); in practice, these corrections make little difference for analyses made in this paper. We use uniform weights, $w = 1$, to compute the redshift space correlation function, $\xi(s)$. The mean densities $n_R$ and $n_D$ are merely the total number of real galaxies and random galaxies, respectively, divided by the volume (accounting for the excluded regions) of the volume-limited subsamples.

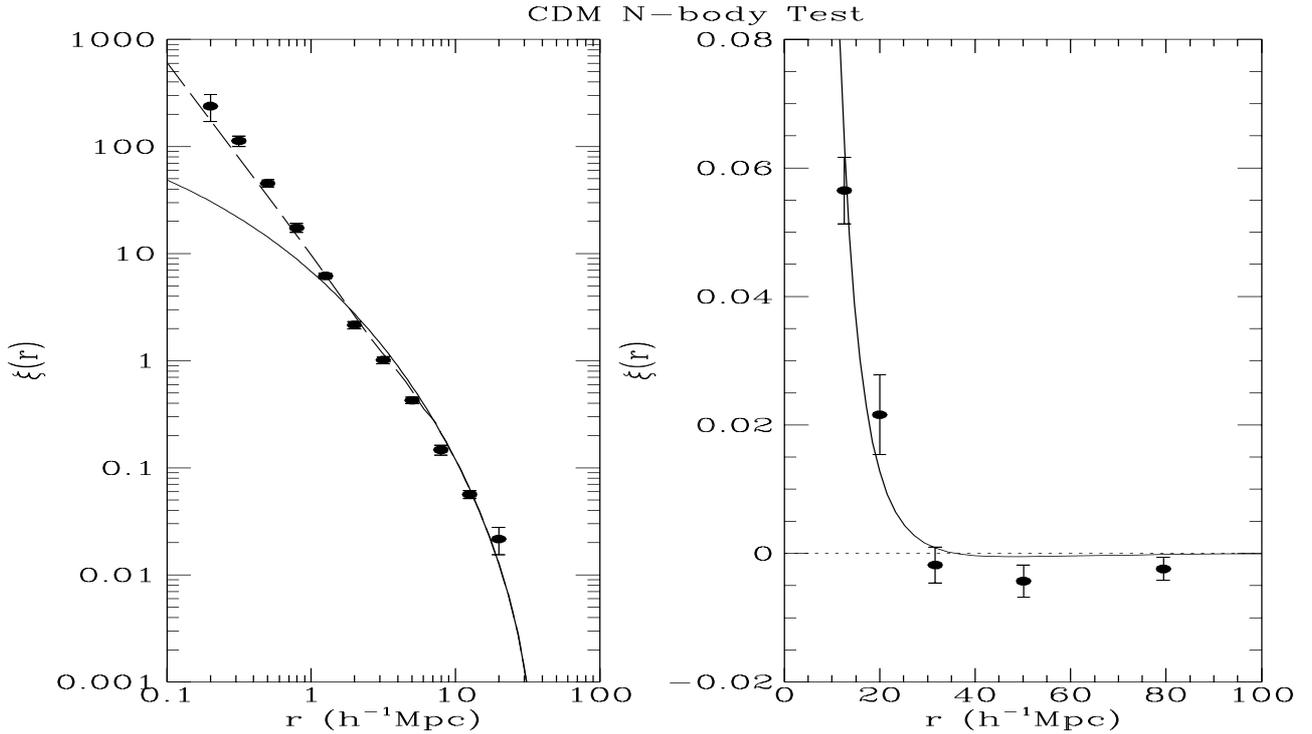

**Figure 4.** Test of the optimally weighted $\xi(r)$. The triangles represent the mean optimally weighted $\xi(r)$ for ten mock *IRAS* catalogs; the error bars are the standard deviation of the mean. The solid line is the linear theory prediction while the dashed line is the full nonlinear $\xi(r)$.

We extracted four different subsamples from the 1.2 Jy survey, which were volume-limited at 60, 80, 100, and 120 $h^{-1}$ Mpc and contained 876, 766, 704, and 575 galaxies, respectively. For each subsample, we created a random catalog of 5000 points uniformly distributed in the volume of space outside the *IRAS* catalog's excluded zones. Error estimates were made from 50 bootstrap resamplings of the data (cf., § 2.2 above). The resulting correlation functions, $\xi(s)$, and the associated bootstrap errors are shown in Figure 5.

We have performed power law fits to the correlation functions, $\xi(s) = (s_0/s)^\gamma$, for separations $1 < s < 13 \ h^{-1}$ Mpc. The results are shown in Table 1. The values of $\xi(s)$ at different separations are not statistically independent; Appendix A describes a method to fit power laws taking this covariance into account. The 68, 90, and 99% confidence intervals for the $(s_0, \gamma)$ for each volume limit are shown in Figure 6. Table 1 also lists the goodness of fit, $Q$, for the fits under the assumption that the residuals are Gaussian. The high values of $Q$ (or equivalently the low values of $\chi^2$) suggest that the bootstrap errors used in the fitting procedure are probably overestimates of the true errors by roughly a factor of two, consistent with the lower panel of Figure 2. All the volume limited subsamples are consistent with $s_0 \sim 4.2 \ h^{-1}$ Mpc and $\gamma \sim 1.4$. The error limits rapidly increase with the volume limit, as the sampling density decreases.

We see no evidence for a correlation between $s_0$ and the depth of the sample in the 1.2 Jy survey as might be expected if the galaxy distribution on these scales were a pure fractal (e.g., Pietronero 1987). The same conclusion was reached by Davis *et al.* (1988) in an analysis of volume limited subsamples of all galaxies in the current sample with fluxes greater than 1.936 Jy (Strauss *et al.* 1990, 1992a). Bouchet *et al.* (1993) did in fact find a weak correlation between clustering strength in the *IRAS* 1.2 Jy sample and the volume limiting radius, which they interpreted as a luminosity effect. However, it was largely limited to smaller volumes than those probed in Figure 5.

### 3.2 Optimal Weighting: $\xi(s)$ on Large Scales

In an effort to make full use of the data, we have also computed $\xi(s)$ from the entire flux-limited survey using the minimum variance weights described in § 2.1. We use all the data with redshifts in the interval 500 km s$^{-1}$ $< s <$ 30,000 km s$^{-1}$ and a random background catalog containing 50,000 points. The resulting correlation function is shown in Figure 7 and listed in Table 2. The errors shown in Figure 7 are the bootstrap errors over 25 resamplings of the data (cf., § 2.2). The correlation function is described quite well by a power law correlation function on scales $\lesssim 20 \ h^{-1}$ Mpc. Using the method in Appendix

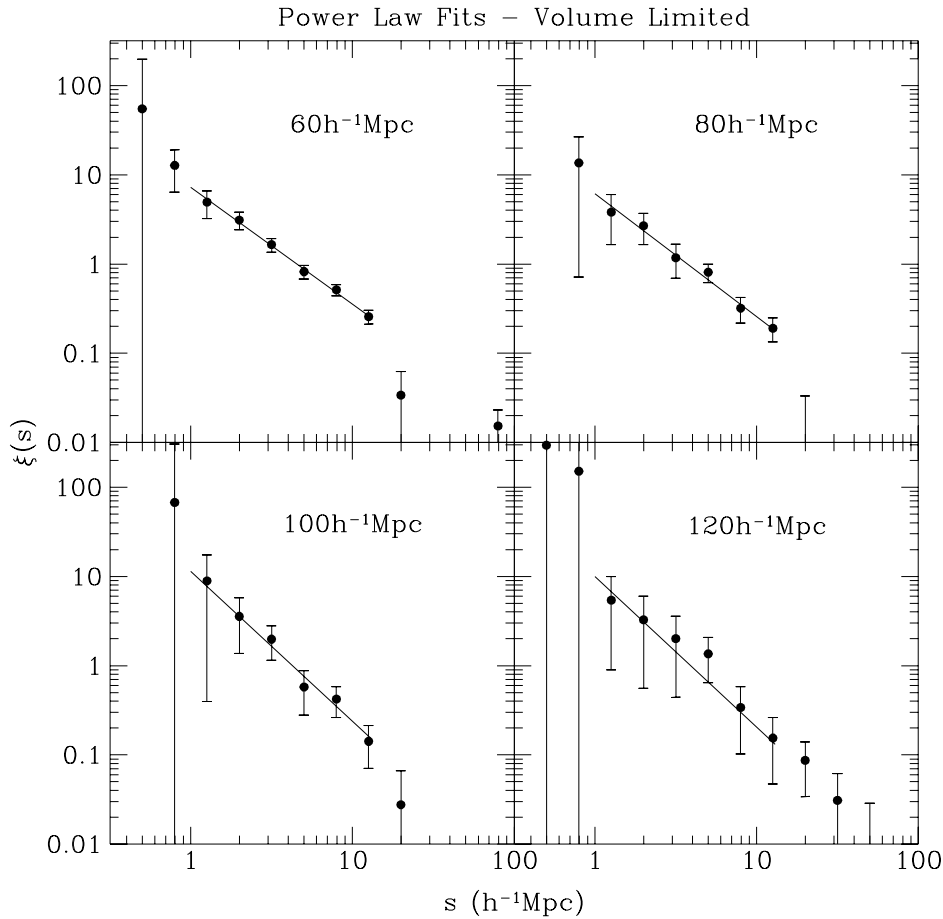

**Figure 5.** Redshift space correlation function for volume limited subsamples. The solid lines are derived by fitting $\xi(s)$ in the range $1 < s < 13\ h^{-1}$ Mpc. Error bars are derived from 25 bootstrap resamplings of the data.

A, we derive a best fit correlation function for $1 < s < 13\ h^{-1}$ Mpc of $s_o = 4.53^{+0.21}_{-0.22}$ and $\gamma = 1.28^{+0.06}_{-0.02}$ (where the errors are 1 $\sigma$, cf., Table 1). On larger scales, $\xi(s)$ begins to deviate from a simple power law. Bouchet et al. (1993) found a steeper power law on small scales, $\gamma = 1.59$. As they discuss, this discrepancy is probably due to a weak luminosity segregation effect: the optimally weighted correlation function on the smallest scales is largely determined by nearby pairs of low luminosity, which show weaker correlations than do the bulk of the galaxies, slightly decreasing the derived $\gamma$.

The confidence intervals for $(s_o, \gamma)$ using the minimum variance weighting scheme are shown in the right-hand panel of Figure 8. The constraints on the model parameters are much tighter than for the volume limited subsamples (cf., Figure 4), due both to the larger number of galaxies and to the use of the minimum variance weights.

Our estimates of $s_o$ and $\gamma$ in the 1.2 Jy sample, which supersede those for the 1.936 Jy sample in Strauss et al. (1992a), are in good agreement with previous results in the literature. Babul & Postman (1989) found $s_o = 3.94^{+1.23}_{-0.91}\ h^{-1}$ Mpc and $\gamma = 1.28 \pm 0.3$ using a volume limited sample of 174 *IRAS* galaxies. Davis et al. (1988) found $s_o \sim 5\ h^{-1}$ Mpc and $\gamma \sim 1.7$ using volume limited samples in the original 1.936 Jy redshift survey. Our value of $s_o$ is discrepant with that found by Moore et al. (1993), who found $s_o = 3.53 \pm 0.35\ h^{-1}$ Mpc and $\gamma = 1.23 \pm 0.17$, in their analysis of the QDOT "1 in 6" *IRAS* redshift survey. However, Moore et al. used an indirect method for determining $\xi(s)$ by modeling the effect of redshift distortions on the real space correlation function, $\xi(r)$, of the QDOT sample as determined by Saunders et al. (1992). This approach relies on the proper modeling of the peculiar velocity field and redshift errors; these points are addressed in the second paper of this series (Fisher et al. 1993b).

In Figure 9, we compare the *IRAS* 1.2 Jy $\xi(s)$ with a variety of linear theory models. The linear theory models have been normalized so that the variance in a sphere of radius $8\ h^{-1}$ Mpc matches the observed variance of *IRAS* galaxies when measured in redshift space, $\sigma_8 = 0.8$ (cf., § 5 below). Figure 9 should not be used as a basis for quantitative comparisons between different models, for at least three reasons: the models do not include non-linear effects, redshift distortions change the shape of $\xi(s)$ in the transition regime where $\xi(s) \approx 1$, and the bootstrap errors on the derived $\xi(s)$ are probably overestimated (Figure 2). A proper comparison requires an extensive Monte-Carlo analysis based on $N$-body simulations (see Fisher *et al.* 1993a for such an analysis for the *IRAS* 1.2 Jy power spectrum).

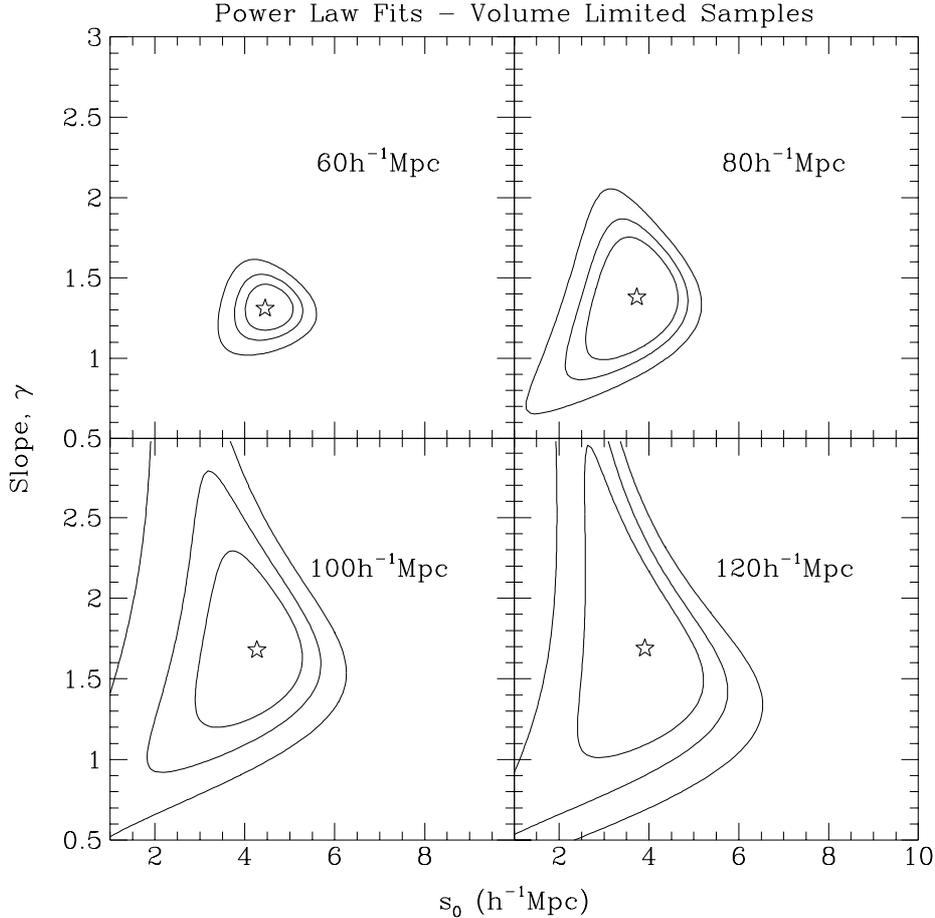

**Figure 6.** Confidence intervals for volume-limited samples. The power law models were fit in the range $1 < s < 13\ h^{-1}$Mpc. Contours correspond to $\Delta\chi^2$ of the 68, 90, and 99% confidence levels (cf., Appendix A). Overestimation of errors by the bootstrap process may mean that these confidence intervals are too wide.

The solid curve in Figure 9 is the standard CDM correlation function ($\Omega = 1$, $h = 0.5$, $\Omega_\Lambda = 0$). We see evidence for correlations in excess of those predicted by the standard CDM model on scales $s \gtrsim 15\ h^{-1}$Mpc. The discrepancy is most significant at the 20 $h^{-1}$Mpc point where the *IRAS* $\xi(s) = 0.11 \pm .036$ and the standard CDM ($\sigma_8 = 0.8$) model predicts $\xi(s) = 0.021$. The CDM correlation function goes negative at 35 $h^{-1}$Mpc; unfortunately the errors in the *IRAS* correlation function are rapidly increasing at this scale and the statistical significance of the discrepancy with the standard CDM model is not high (at 32 $h^{-1}$Mpc, the *IRAS* $\xi(s)$ is only $\sim 1\sigma$ above zero).

The dotted curve in Figure 9 shows the CDM spectrum for $\Omega = 0.2$ and $h = 1.0$ and with a cosmological vacuum energy density of $\Omega_\Lambda = 0.8$. This model has significantly more power than does standard CDM; Efstathiou *et al.* (1990a) have shown that this model nicely fits the angular correlation function from the APM galaxy survey. This model provides a much better fit to the *IRAS* 1.2 Jy $\xi(s)$ than does standard CDM, as it does for the *IRAS* power spectrum (Fisher *et al.* 1993a).

We also show the correlation function for three models based on structure formed by topological vacuum defects in the early universe. The short dashed-dotted curve corresponds to a CDM universe seeded with cosmic strings (Albrecht & Stebbins 1992a). This model produces excessive small scale power and insufficient correlations on large scales to be consistent with the *IRAS* results. The short dashed curve in Figure 9 shows the correlation function for another string model; in this case the mass density of the universe is dominated by Hot Dark Matter (HDM) in the form of a massive neutrino (Albrecht & Stebbins 1992b). The free streaming HDM tends to suppress small scale fluctuations. Consequently, when normalized to galaxy scales, the effect of HDM is to boost the amount of large scale power relative to the CDM+strings model. The HDM+strings $\xi(s)$, however, still lacks sufficient amplitude on scales $\gtrsim 10\ h^{-1}$Mpc to be consistent with the *IRAS* $\xi(s)$.

The long dashed-dotted curve in Figure 9 corresponds to a CDM universe seeded by textures with $h = 0.5$ and $\Omega = 1$ (cf., Gooding, Spergel, & Turok 1991). The curve shown was derived from the power spectrum of Turok (1991) as quoted in Peacock (1991). The model fits the *IRAS* $\xi(s)$ quite well, and looks very similar to the $\Omega h = 0.2$ CDM model discussed above. Unfortunately, textures are likely to produce microwave background temperature fluctuations which are inconsistent with the recent COBE measurements (Smoot *et al.* 1992; Wright *et al.* 1992) unless the galaxy distribution is characterized by a high bias parameter, $b = 2.0 \pm 0.5 h^{-1}$ (Pen, Spergel, & Turok 1993).

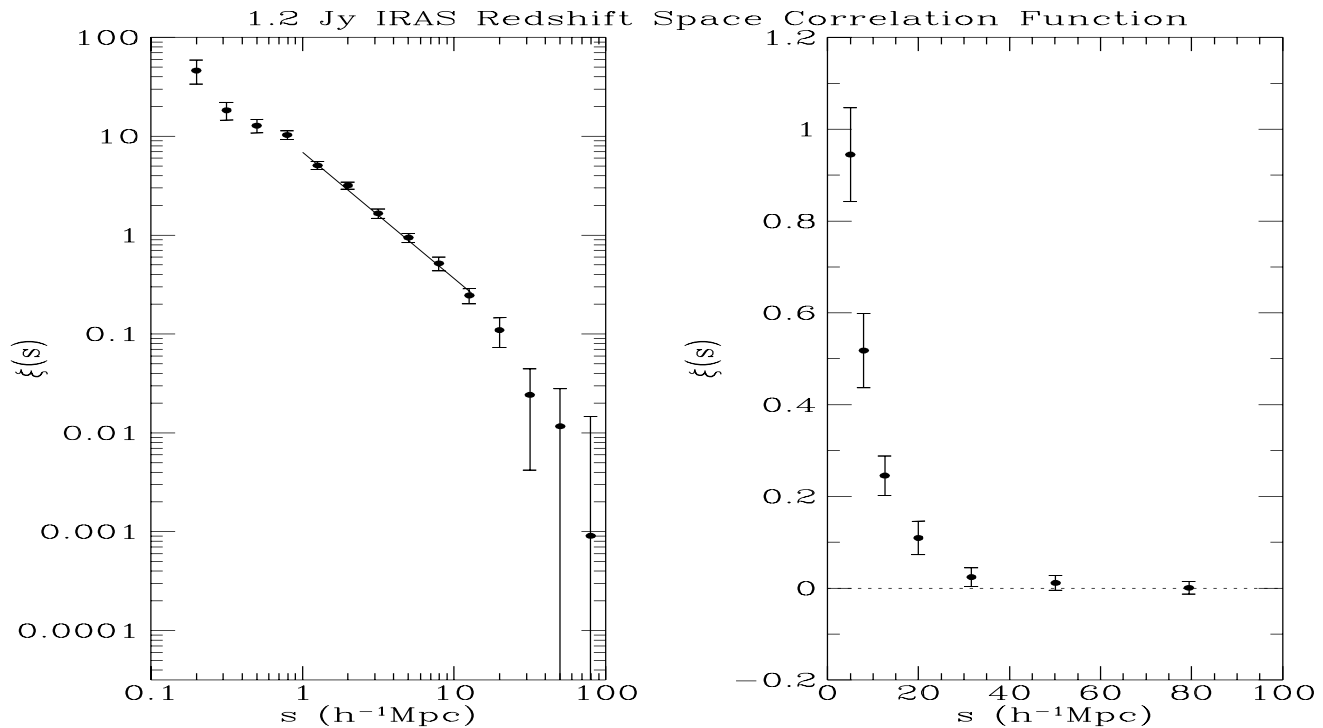

**Figure 7.** The *IRAS* 1.2 Jy redshift space correlation function determined using the minimum variance weighting scheme. Bootstrap error bars are shown. The fit shown is done to $1 < s < 13\ h^{-1}$Mpc.

The long dashed curve in Figure 9 shows the correlation function for a hybrid universe containing both CDM and HDM. The correlation function was computed from the power spectrum given in Holtzman (1989) corresponding to a universe with 70% of its mass in the form of CDM and 30% in the form of a massive neutrino. The curve shown is for a Hubble constant, $h = 0.5$. This model seems to fit the *IRAS* $\xi(s)$ reasonably well; it has also been noted (Wright *et al.* 1992) that this model is in good agreement with the COBE measurements, and has the potential to solve a number of other cosmological puzzles (Taylor & Rowan-Robinson 1992; Davis, Summers, & Schlegel 1992; Schaefer & Shafi 1992; Klypin *et al.* 1992; Cen & Ostriker 1993).

## 4 REDSHIFT SPACE DISTORTIONS

### 4.1 The Real Space Correlation Function

Up to now, we have been measuring the correlation function in redshift space, $\xi(s)$. The redshift space correlation function differs from the correlation function in real space on both small and large scales. On small scales, internal random motions in bound groups of galaxies lead to structures elongated along the line of sight in redshift space (the so-called "fingers of God"). On large scales, the peculiar velocities of galaxies also lead to a distortion of the correlation function, now due to coherent motions of galaxies rather than the randomized velocities characteristic of virialized groups. Imagine an overdense region of space which induces the coherent infall of nearby galaxies. When viewed in redshift space, galaxies in the foreground of the overdensity will have redshifts in excess of their distances because their velocities will be flowing away from the observer and towards the overdensity. Conversely, objects behind the overdensity will have peculiar velocities directed towards the observer and will thus have redshifts which are less than their actual distances. The combined effect is to compress the structure along the observer's line of sight (cf., Kaiser (1987) for a quantitative discussion of this effect within the context of linear theory).

We quantify the effects of the redshift space distortions by computing the correlation function as a function of separations parallel and perpendicular to the observer's line of sight. Explicitly, given a pair of galaxies with redshift positions, $\mathbf{v}_1$ and $\mathbf{v}_2$, we can define both a separation in redshift space and observer's line of sight by $\mathbf{s} = \mathbf{v}_1 - \mathbf{v}_2$ and $\mathbf{l} = \frac{1}{2}(\mathbf{v}_1 + \mathbf{v}_2)$ respectively. We then define separations which are parallel ($\pi$) and perpendicular ($r_p$) to the line of sight,

$$\pi = \frac{\mathbf{s} \cdot \mathbf{l}}{|\mathbf{l}|} , \tag{10}$$

$$r_p^2 = \mathbf{s} \cdot \mathbf{s} - \pi^2 . \tag{11}$$

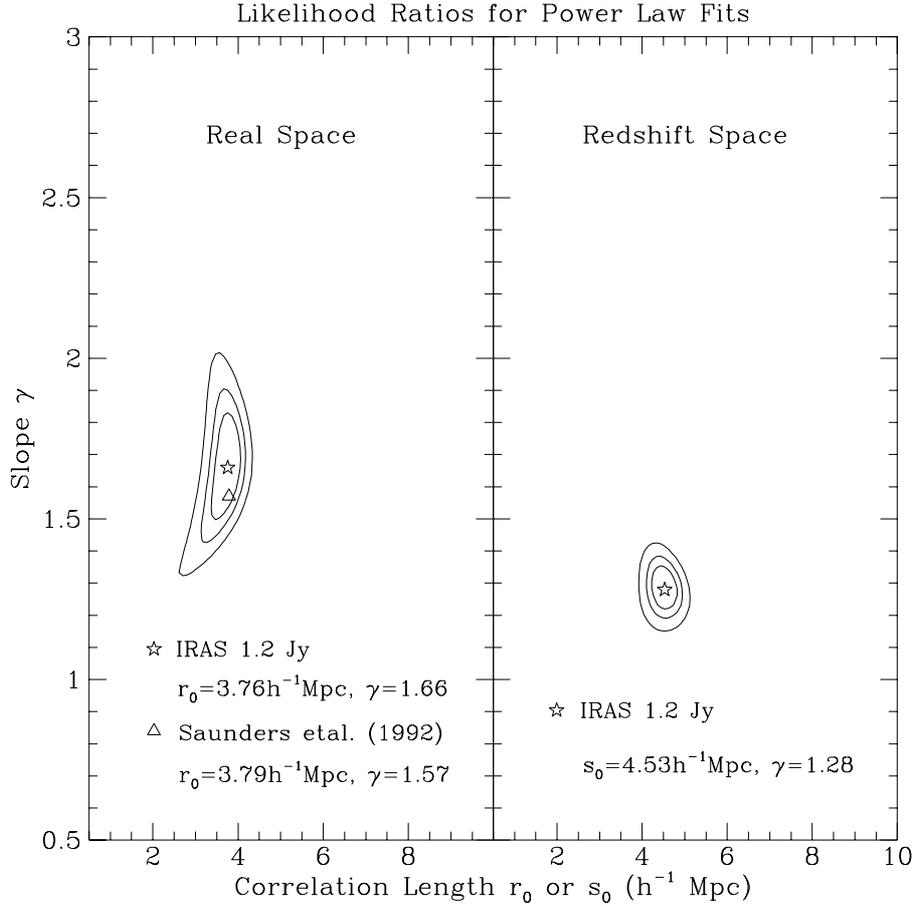

**Figure 8.** 68, 90, and 99% confidence intervals for optimal weighted flux limited samples. The left hand panel is the correlation function in real space, $\xi(r)$, derived from the deprojection of $\xi(r_p, \pi)$. The right hand panel corresponds to the redshift space correlation function, $\xi(s)$. Overestimation of errors by the bootstrap process may mean that these confidence intervals are too wide.

With these new variables, we can write down a generalized version of the estimator for the correlation function,

$$\xi(r_p, \pi) = \frac{N_{DD}(r_p, \pi)}{N_{DR}(r_p, \pi)} \frac{n_R}{n_D} - 1 , \tag{12}$$

where $N_{DD}(r_p, \pi)$ and $N_{DR}(r_p, \pi)$ refer to data-data and data-random pairs with separations $\pi$ and $r_p$ respectively.

Figure 10 shows $\xi(r_p, \pi)$ for the full 1.2 Jy *IRAS* sample. The $\xi(r_p, \pi)$ in Figure 10 was computed using all the galaxies in the flux-limited sample in the redshift range $500 < cz < 30,000$ km s$^{-1}$ using the minimum variance weights given in Equation 4, with bin widths and separations of 1 $h^{-1}$Mpc. The contours are linear with $\Delta\xi(r_p, \pi) = 0.1$ for $\xi(r_p, \pi) < 1$ and logarithmic (0.1 in dex) for $\xi(r_p, \pi) > 1$; the heavy contour denotes $\xi(r_p, \pi) = 1$ while the dashed curves denotes negative contours, $\xi(r_p, \pi) < 0$. The dotted concentric curves correspond to the redshift space correlation function, $\xi(s)$, at $\xi(s) = 1.0$, 0.5, and 0.25. For graphical clarity, the map has been twice smoothed by a 1-2-1 boxcar in each direction. The small scale distortion discussed above is evident as the stretching of the $\xi(r_p, \pi)$ contours along the $\pi$ direction for values of $r_p \lesssim 2\ h^{-1}$Mpc. The weak compression of the contours along the $\pi$ axis for $r_p \gtrsim 5\ h^{-1}$Mpc is the signature of the large scale redshift distortion.

In a companion paper (Fisher *et al.* 1993*b*) we present a detailed study of the redshift distortions using $\xi(r_p, \pi)$. In this paper, however, we limit our analysis of $\xi(r_p, \pi)$ to a recovery of the correlation function in real space. Following Davis &

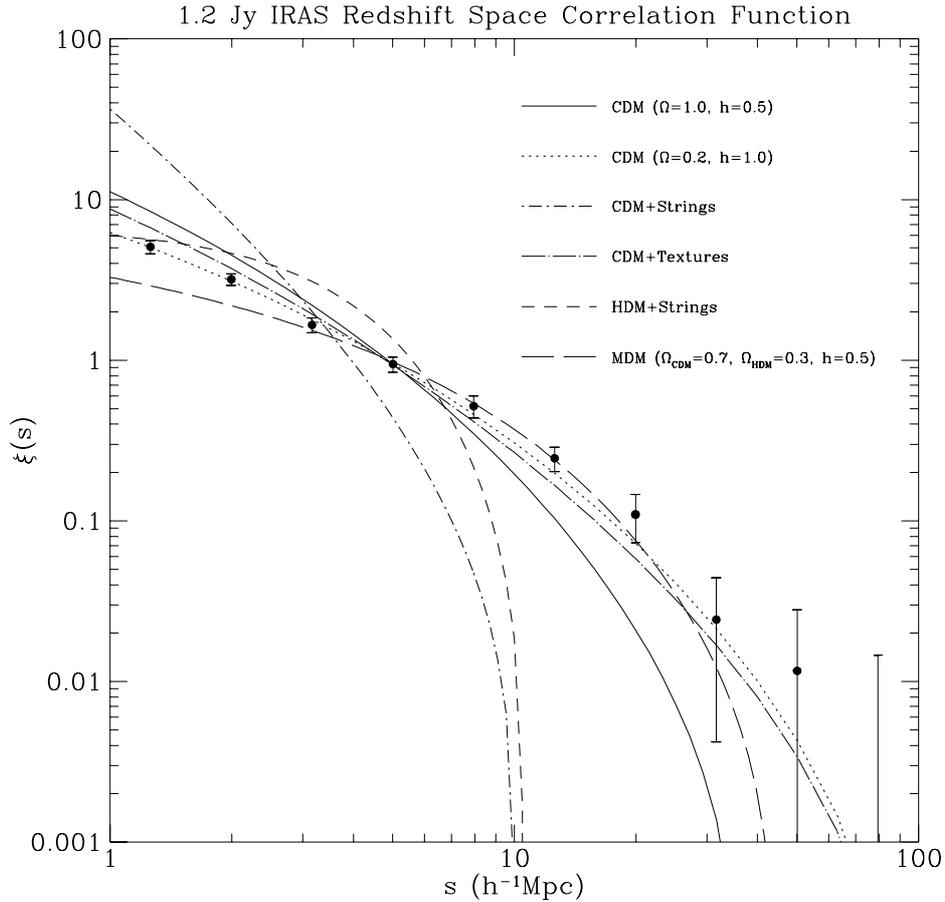

**Figure 9.** Comparison of $IRAS$ $\xi(s)$ with various linear theory models described in text. Models have been normalized to the variance in redshift space of $IRAS$ galaxies in a sphere of radius 8 $h^{-1}$ Mpc, $\sigma_8 = 0.80$ (cf., § 5). The solid line is the correlation function of the standard ($\Omega h = 0.5$) CDM model.

Peebles (1983), we define a projected function which is unaffected by redshift distortions,

$$w_p(r_p) \;=\; 2 \int_0^\infty d\pi\, \xi(r_p, \pi) \tag{13}$$

$$\;=\; 2 \int_0^\infty dy\, \xi\left[(r_p^2 + y^2)^{1/2}\right] \quad . \tag{14}$$

The integrand in the second expression for $w_p(r_p)$ is the correlation function in real space. If we model $\xi(r)$ as a power law, the integral for $w_p(r_p)$ can be performed analytically to give

$$w_p(r_p) = r_p \left(\frac{r_o}{r_p}\right)^\gamma \frac{\Gamma\left(\frac{1}{2}\right)\Gamma\left(\frac{\gamma-1}{2}\right)}{\Gamma\left(\frac{\gamma}{2}\right)} \quad , \tag{15}$$

where $\Gamma(x)$ is the usual Gamma function. We have constructed $w_p(r_p)$ from the map of $\xi(r_p, \pi)$ given in Figure 10 by numerically performing the integration in Equation 13; the resulting $w_p(r_p)$ is shown in Figure 11. We then fit a power law to $w_p(r_p)$ using Equation 15 and the method described in Appendix A. We use a set of 25 bootstrap resamplings to provide an error estimate for the fit. The recovered real space correlation function is ($1\sigma$ errors)

$$\xi(r) = \left(\frac{r}{3.76^{+0.20}_{-0.23} h^{-1}\,\mathrm{Mpc}}\right)^{1.66^{+0.12}_{-0.09}} \quad . \tag{16}$$

The confidence intervals for $(r_o, \gamma)$ are shown in the left hand panel of Figure 8. The power law $w_p(r_p)$ for the best fit parameters is shown as the solid line in Figure 11, where the fit is confined to the range $1 < r_p < 13~h^{-1}$Mpc. On larger scales $w(r_p)$ drops faster than the power law model.

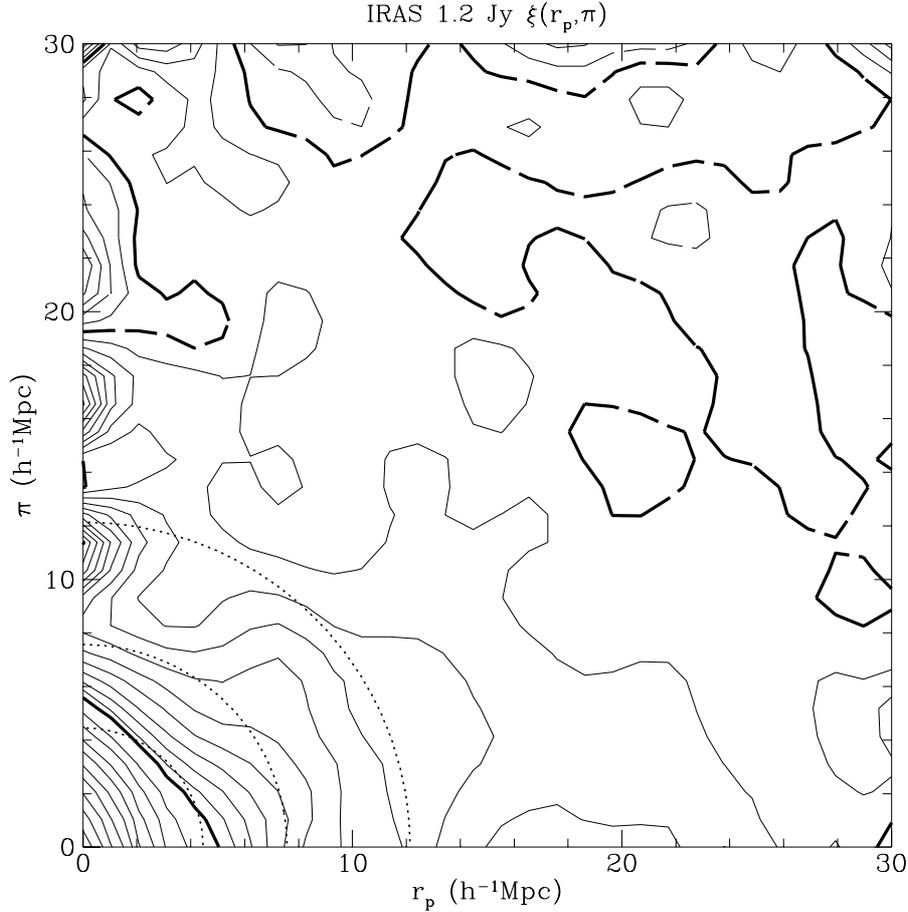

**Figure 10.** $\xi(r_p, \pi)$ for the 1.2 Jy *IRAS* sample. The contours are in steps of $\Delta\xi(r_p, \pi) = 0.1$ for $\xi(r_p, \pi) < 1$ and logarithmic (0.1 dex) for $\xi(r_p, \pi) > 1$. The heavy solid contour is at $\xi(r_p, \pi) = 1$. Dashed contours represent $\xi(r_p, \pi) < 0$ with the heavy dashed contour at $\xi(r_p, \pi) = 0$. The concentric dotted lines are the angle averaged redshift space correlation function, $\xi(s)$ at $\xi(s)=1.0$, 0.5, and 0.25.

Our inferred $\xi(r)$ is in excellent agreement with that of Saunders *et al.* (1992) (Equation 7), who obtained $\xi(r)$ for *IRAS* galaxies using a technique based on the cross-correlation between the QDOT redshift survey of *IRAS* galaxies (Lawrence *et al.* 1993) and its parent 2-D catalogue, the QIGC survey (Rowan-Robinson *et al.* 1991).

The *IRAS* $\xi(r)$ is also consistent with that measured for optically selected spiral galaxies; Davis & Geller (1976) found $r_o = 3.6~h^{-1}$Mpc and $\gamma = 1.69$ for a sample of spiral galaxies selected from the Nilson (1973) catalog, while Giovanelli *et al.* (1986) inferred $\gamma = 1.69 \pm 0.03$ from the $w(\theta)$ measured for similar set of spirals. The slope determined here is in agreement with that of the optically selected APM galaxies $\gamma = 1.660 \pm 0.014$ (Maddox *et al.* 1988). The difference in the optical and *IRAS* bias factors may be partly explained by the underrepresentation of *IRAS* galaxies in cluster cores; however, Strauss *et al.* (1992a) show that optical galaxies are more strongly clustered than *IRAS* galaxies outside of clusters.

## 5 COUNTS

A very useful moment of the correlation function is given by the separation averaged value of $\xi(r)$ in a volume, $V$,

$$\sigma^2 = \frac{1}{V^2} \int_V dV_1 \, dV_2 \, \xi(|\mathbf{r}_1 - \mathbf{r}_2|) \quad . \tag{17}$$

$\sigma^2$ has the simple physical interpretation as the variance of counts (in excess of Poisson) in the volume, i.e.,

$$\langle (N - nV)^2 \rangle = nV + n^2 V^2 \sigma^2 \quad , \tag{18}$$

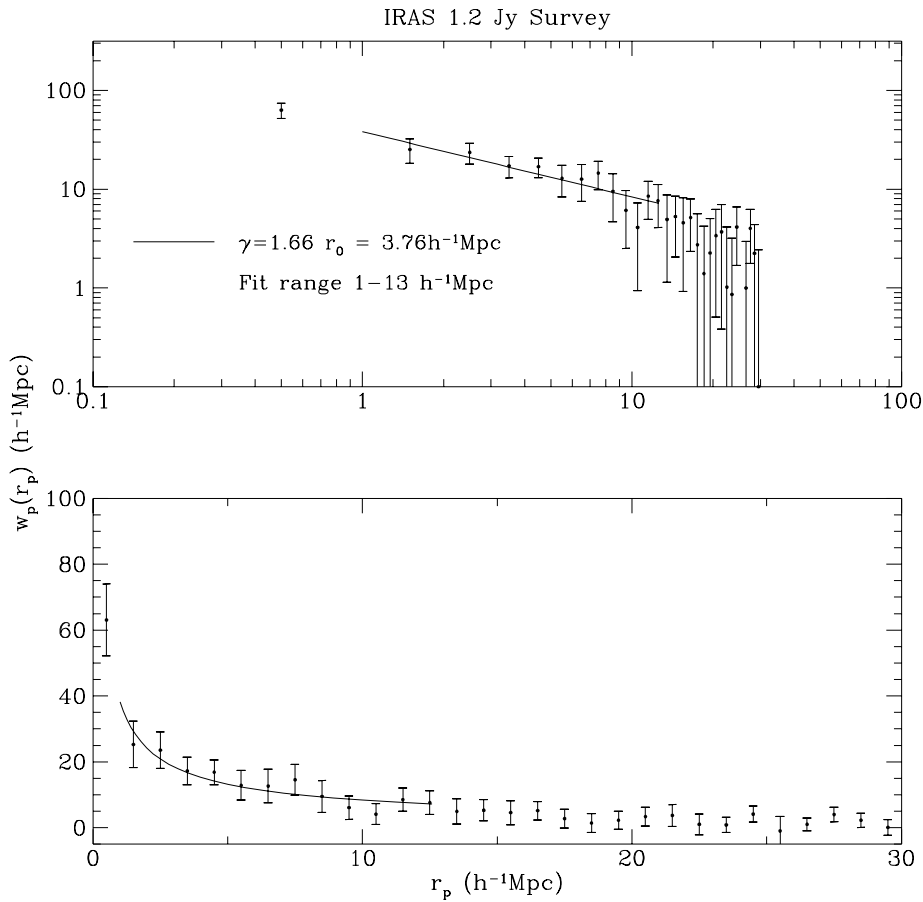

**Figure 11.** The projected $\xi(r_p, \pi)$ function, $w_p(r_p)$ for the 1.2 Jy *IRAS* sample. The top and bottom panels show $w_p(r_p)$ on logarithmic and linear scales respectively. Error bars are derived from 25 bootstrap resamplings of the data. The solid curve in both panels represents the best fit power law model (fit for $1 < r_p < 13\ h^{-1}$ Mpc) of $r_o = 3.76\ h^{-1}$ Mpc and $\gamma = 1.66$.

where $nV$ is the mean number of galaxies in the volume $V$ (cf., Peebles 1980, § 36.6). The first term on the right in Equation 18 is the usual Poisson "shot noise" contribution.

For a power law correlation function and a spherical region of radius $R$, the integral in Equation 17 can be performed analytically to yield

$$\sigma^2_{sphere}(R) = \frac{72(r_o/R)^\gamma}{[2^\gamma(3-\gamma)(4-\gamma)(6-\gamma)]} \qquad (19)$$

(cf., Peebles 1980, § 59.3). Theoretical models for the power spectrum are often normalized to the galaxy distribution by matching the variance in 8 $h^{-1}$ Mpc spheres, $\sigma_8 \equiv \sigma(R = 8\ h^{-1}\text{Mpc})$, to the observed value. The values of $\sigma_8$ computed using Equation 19 for the power law fits to the 1.2 Jy *IRAS* correlation function are given in Table 1. In particular, the real space correlation function derived from $\xi(r_p, \pi)$ gives $\sigma_8 = 0.69 \pm 0.04$. The $\sigma_8$ in redshift space from the minimum variance weighted $\xi(s)$ is $\sigma_8^{red} = 0.8 \pm 0.03$, which is slightly greater than $\sigma_8$ determined in real space due to the effect of redshift space distortions (e.g., Kaiser 1987). One might hope that the ratio of the redshift space and real space variances can be used to estimate the quantity $\Omega^{0.6}/b$ using the results of Kaiser (1987). Unfortunately, $\sigma_8$ is dominated by scales which are not in the linear regime, so the linear theory distortion factor given in Kaiser (1987) is not strictly valid. Moreover, the real and redshift space *IRAS* correlation functions have different slopes and consequently their ratio is a function of scale. It would be interesting to determine the ratio of the cumulative real and redshift correlation functions, i.e., a ratio of the real and redshift space $J_3(r)$ integrals, on scales of $10 - 20\ h^{-1}$ Mpc. Unfortunately, the clean inversion of $w_p(r_p)$ to obtain $\xi(r)$ requires an *a priori* model for $\xi(r)$ and on scales $r \gtrsim 15\ h^{-1}$ Mpc, $w_p(r_p)$ begins to deviate from a simple power law.

The real space *IRAS* correlation function is very similar in slope to the result of Maddox et al. (1988) for the APM survey, suggesting that the ratio of the *IRAS* to optical correlation functions is at most a slowly varying function of scale. Thus on scales $\gtrsim r_o$, the ratio of the real space variances from the optical and *IRAS* samples gives a direct estimate of the optical to *IRAS* bias ratios. For the real space *IRAS* $\xi(r)$ listed in Table 1 and an optical correlation function with $(r_0 = 5.4 \pm 0.3~h^{-1}\mathrm{Mpc},~\gamma = 1.77 \pm 0.04)$ (Davis & Peebles 1983), the variances given by Equation 19 are

$$\sigma_8^{IRAS} = 0.69 \pm 0.04~, \tag{20}$$

$$\sigma_8^{Optical} = 0.95 \pm 0.06~, \tag{21}$$

$$\left(\frac{b_O}{b_I}\right)_{\sigma_8} = 1.38 \pm 0.12~.$$

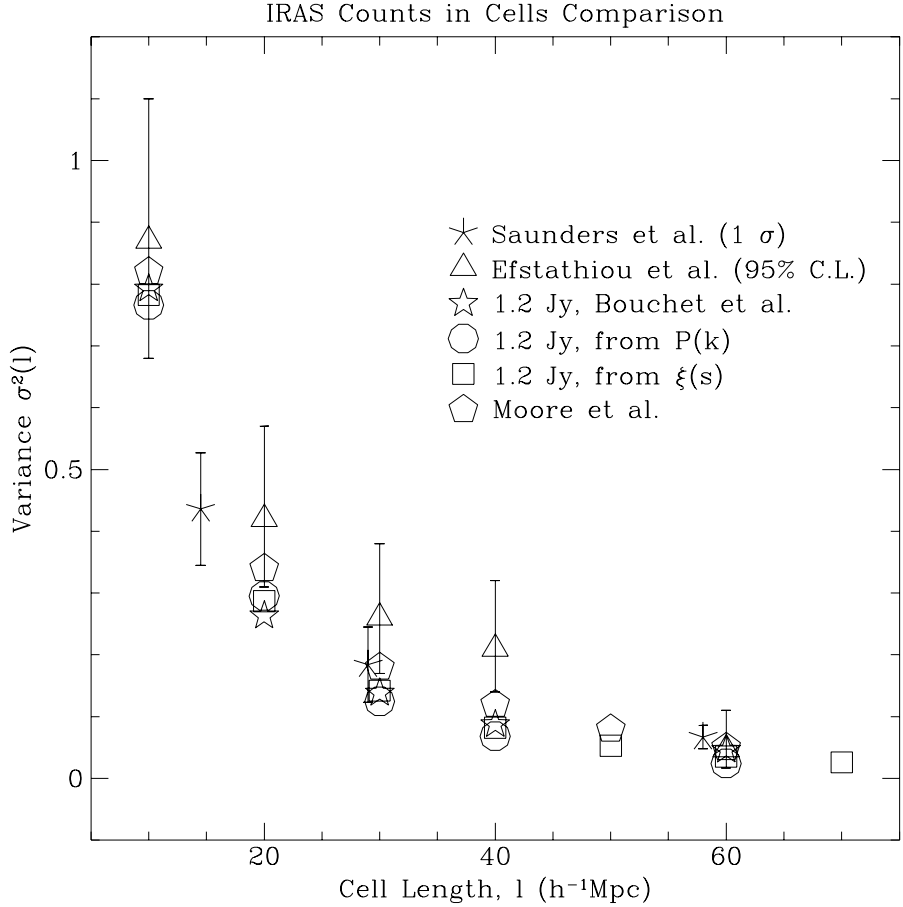

**Figure 12.** Comparison of the variances in cubical cells for various results in the literature. The errors on the Saunders *et al.* (1991) points are 1σ, while the errors on the Efstathiou *et al.* points represent 95% confidence intervals. The errors on the variances derived from the *IRAS* survey are given in Table 4.

The $\sigma_8$ subscript is given on the bias ratio in the above equation to emphasize the fact that it is determined on scales which could be contaminated by nonlinear evolution. The optical to *IRAS* bias ratio given above is consistent with that reported by Lahav *et al.* (1990), $b_O/b_I \approx 1.4$–2.3, from an analysis of the *IRAS* angular correlation function, and by Strauss *et al.* (1992a) from an analysis of the correlation functions of optical and *IRAS* galaxies in *redshift* space.

We can also compute the variances directly from the redshift space correlation function by integrating a spline fit to $\xi(s)$ in Equation 17. Table 4 gives the results for $\sigma^2(l)$, defined as the variance of counts in cubical cells with sides of length $l~h^{-1}$Mpc. The error estimates for $\sigma^2(l)$ given in Table 4 are one half the difference of the variances obtained with all the points in the spline fit to $\xi(s)$ perturbed up and down, respectively, by their 1σ bootstrap errors (cf., § 2.2). These errors are undoubtedly overestimates since the bootstrap errors overestimate the true errors (cf., Figure 2) and because the individual bins in $\xi(s)$ are not 100% correlated. Therefore, one should exercise caution in drawing rigorous conclusions from these error estimates.

In Figure 12, we compare the values of $\sigma^2(l)$ inferred from our correlation function with previous results in the literature as well as with other determinations from the 1.2 Jy survey. The precise numbers and error estimates for the data in Figure 12 are given in Table 4. Our results for $\sigma^2(l)$ determined from $\xi(s)$ can be directly compared with the results of Moore *et* al. (1993) who performed a similar numerical integration of the QDOT redshift correlation function to obtain the variance in cubical cells. The Moore et al. results are shown as the pentagons in Figure 12; their results are consistently higher than those determined from the 1.2 Jy survey but the 1.2 Jy $\sigma^2(l)$ are within their quoted 1 $\sigma$ errors for all reported cell sizes (cf., Table 4). The open squares in Figure 12 represent the values of $\sigma^2(l)$ derived here from the *IRAS* 1.2 Jy $\xi(s)$. The triangles in Figure 12 show the variances in cubical cells in volume limited subsamples of the QDOT survey (Efstathiou et al. 1990b; hereafter E90); the error bars on the triangles shown in Figure 12 are their quoted 95% confidence intervals. The closed stars in Figure 12 show an estimate of the variances in the QDOT survey derived from a Gaussian smoothed galaxy density field (Saunders et al. 1991) along with the corresponding 2 $\sigma$ errors.

Also shown as the open stars in Figure 12 (and in Table 4) are the variances derived directly from the 1.2 Jy survey by computing the number counts in spheres randomly placed within volume limited subsamples (Bouchet et al. 1993). We have used the approximation given in Bouchet et al. to scale their variances to the corresponding variances in cubical cells. In Fisher et al. (1993a), we derive the Fourier conjugate of the correlation function, the power spectrum, $P(k)$, for the 1.2 Jy *IRAS* survey. The variances derived from the power spectrum are listed in Table 4 and are shown as the open circles in Figure 12.

The agreement between the 1.2 Jy variances derived from $\xi(s)$, $P(k)$, and direct counts is striking. The error-weighted mean variances over the determinations from the 1.2 Jy sample are shown in Table 4 .

The 1.2 Jy *IRAS* variances, however, appear discrepant (at greater than the 95% confidence level) with the results of E90 for the 20, 30, and 40 $h^{-1}$ Mpc cell sizes. It has been suggested that the QDOT estimate of the variance at 40 $h^{-1}$ Mpc could be a statistical fluke (Park 1991); the disagreement between the two samples on smaller scales, however, seems to suggest an underestimation of the errors in the QDOT $\sigma^2(l)$ by E90.

We have begun a collaboration with George Efstathiou to compare the 1.2 Jy and QDOT survey density fields point by point to resolve the discrepancies found in Figure 12; the results of this work will be presented at a later date.

## 6 CONCLUSIONS

Our results for the redshift space correlation function, $\xi(s)$, in the 1.2 Jy *IRAS* survey for scales $s \lesssim 20$ $h^{-1}$ Mpc are well described by a power law with parameters $s_0 = 4.5$, $\gamma = 1.3$ (optimal weighting), in good agreement with those found in previous analyses of *IRAS* surveys. On larger scales $s \gtrsim 20$ $h^{-1}$ Mpc, $\xi(s)$ drops below the power law extrapolation. Despite the inherent problems of normalizing models to the data due to uncertainties in redshift distortions and non-linear effects, the measured $\xi(s)$ appears to favor cosmological models with more power on large scales than predicted by the standard CDM model. This result is in agreement with our analysis of the 1.2 Jy power spectrum (Fisher et al. 1993a) and indicates that the *shape* of the *IRAS* power spectrum (or $\xi(r)$) differs from that predicted by the standard CDM model. A quantitative statement, however, is not possible without accounting for non-linear effects, redshift space distortions, and the possible biases in defining error bars with bootstrap resampling.

We see clear evidence for redshift distortions in our maps of the full correlation function $\xi(r_p, \pi)$: an elongation along $\pi$ on small scales due to nonlinear clustering and a compression along $r_p$ due to the large scale coherent motion of the galaxies. By projecting out the redshift distortions, we have recovered the real space correlation function, $\xi(r)$, which, over the scales over which it can be reliably determined, is in excellent agreement with the previous work of Saunders et al. (1992). The small scale "Finger of God" distortion has been seen clearly in previous investigations of optical catalogs, but this is the first redshift survey with the combination of sampling and volume to reliably detect the flattening of the $\xi(r_p, \pi)$ contours on large scales (cf., Hamilton et al. 1991). The distortions seen in the $\xi(r_p, \pi)$ contain useful information about the nature and coherence of the peculiar velocity field. The results of our modeling of $\xi(r_p, \pi)$ can be found in the second paper of this series (Fisher et al. 1993b).

The real and redshift space correlation functions give accurate values for the overall normalization ($\sigma_8$) of the density fields in both real and redshift space. The real space normalization, $\sigma_8 = 0.69 \pm 0.04$, is 30% lower than the equivalent normalization for optical galaxies ($\sigma_8 = 1$); this difference should be kept in mind when comparing theoretical models to results from the *IRAS* database.

While the two-point correlation functions provide robust and easily interpreted measures of galaxy clustering they discard information, such as phases, about the galaxy distribution. In particular the redshift space correlation function $\xi(s)$ is merely the first moment of the full two dimensional correlation function, $\xi(r_p, \pi)$. By modeling the full structure seen in the $\xi(r_p, \pi)$, one can gain insight about the nature of the peculiar velocity field; this is the goal of the second paper in this series.

**Acknowledgments** We would like to thank Shaun Cole for bringing the method of principal component analysis used in the model fitting procedure to our attention. We thank Simon White, Ben Moore and Robert Lupton for useful discussions. KBF acknowledges a Berkeley Department of Education Fellowship and a SERC postdoctoral fellowship. MAS is supported at the IAS under NSF grant # PHY92-45317, and grants from the W.M. Keck Foundation and the Ambrose Monell Foundation. The work of MD was partly supported by NSF grant AST-8915633 and NASA grant NAG5-1360.

**TABLE 1**
POWER LAW FITS, 1 TO 13 $h^{-1}$Mpc

| Subsample | $r_\circ{}^a$ | $\gamma^a$ | $\chi^2$/d.o.f. | $Q^b$ | $\sigma_8{}^c$ |
|---|---|---|---|---|---|
| Volume-limited ($h^{-1}$Mpc) | | | | | |
| 60 | $4.54^{+0.36}_{-0.34}$ | $1.31^{+0.12}_{-0.07}$ | 0.63/4 | 0.96 | $0.80 \pm 0.05$ |
| 80 | $3.73^{+0.40}_{-0.40}$ | $1.38^{+0.16}_{-0.13}$ | 1.07/4 | 0.90 | $0.70 \pm 0.07$ |
| 100 | $4.27^{+0.66}_{-0.81}$ | $1.68^{+0.36}_{-0.29}$ | 0.89/4 | 0.93 | $0.77 \pm 0.19$ |
| 120 | $3.91^{+0.86}_{-0.86}$ | $1.69^{+0.57}_{-0.40}$ | 1.02/4 | 0.91 | $0.71 \pm 0.27$ |
| Weighted | | | | | |
| *redshift space* | $4.53^{+0.21}_{-0.22}$ | $1.28^{+0.06}_{-0.02}$ | 8.30/8 | 0.40 | $0.80 \pm 0.03$ |
| *real space* | $3.76^{+0.20}_{-0.23}$ | $1.66^{+0.12}_{-0.09}$ | 4.62/8 | 0.80 | $0.69 \pm 0.04$ |

[a] 1 $\sigma$ errors.
[b] Goodness of fit *if* residuals are Gaussian (cf., Appendix A).
[c] Computed from fit using Equation 19.

TABLE 2
OPTIMALLY WEIGHTED $\xi(s)$
as shown in Figure 7

| $s$ ( $h^{-1}$Mpc) | $\xi(s)$ | error[a] |
|---|---|---|
| 0.200 | 46. | 13.0 |
| 0.316 | 18. | 3.8 |
| 0.501 | 13. | 2.0 |
| 0.794 | 10. | 1.0 |
| 1.26 | 5.1 | 0.47 |
| 2.00 | 3.2 | 0.26 |
| 3.16 | 1.7 | 0.18 |
| 5.01 | 0.94 | 0.10 |
| 7.94 | 0.52 | 0.081 |
| 12.6 | 0.24 | 0.043 |
| 20.0 | 0.11 | 0.036 |
| 31.6 | 0.024 | 0.020 |
| 50.1 | 0.012 | 0.016 |
| 79.4 | 0.00091 | 0.014 |
| 126. | -0.0052 | 0.011 |

[a] $1\sigma$ Bootstrap errors.

**TABLE 3**
$w_p(r_p)$
as shown in Figure 11

| $r_p$ ( $h^{-1}$Mpc) | $w_p(r_p)$ | error[a] |
|---|---|---|
| 0.50 | 63.1 | 10.9 |
| 1.5 | 25.3 | 7.06 |
| 2.5 | 23.5 | 5.57 |
| 3.5 | 17.2 | 4.24 |
| 4.5 | 16.8 | 3.78 |
| 5.5 | 12.9 | 4.52 |
| 6.5 | 12.6 | 5.10 |
| 7.5 | 14.5 | 4.68 |
| 8.5 | 9.50 | 4.81 |
| 9.5 | 6.10 | 3.58 |
| 10.5 | 4.10 | 3.16 |
| 11.5 | 8.50 | 3.54 |
| 12.5 | 7.62 | 3.53 |
| 13.5 | 4.94 | 3.80 |
| 14.5 | 5.28 | 3.21 |
| 15.5 | 4.56 | 3.64 |
| 16.5 | 5.16 | 2.82 |
| 17.5 | 2.74 | 2.88 |
| 18.5 | 1.40 | 2.86 |
| 19.5 | 2.26 | 2.77 |
| 20.5 | 3.38 | 2.87 |
| 21.5 | 3.70 | 3.33 |
| 22.5 | 1.02 | 3.14 |
| 23.5 | 0.86 | 2.30 |
| 24.5 | 4.14 | 2.45 |
| 25.5 | -0.98 | 4.43 |
| 26.5 | 1.00 | 1.96 |
| 27.5 | 4.02 | 2.24 |
| 28.5 | 2.24 | 2.15 |
| 29.5 | 0.10 | 2.34 |

[a] $1\sigma$ Bootstrap errors.

**TABLE 4**
COUNTS IN CELLS

| $l$ ( $h^{-1}$ Mpc) | Variance $\sigma^2(l)$ (error) | | | | |
|---|---|---|---|---|---|
| | 10 | 20 | 30 | 40 | 60 |
| 1.2 Jy: (1 $\sigma$ errors) | | | | | |
| $\xi(s)$ | $0.78 \pm 0.10$ | $0.29 \pm 0.05$ | $0.14 \pm 0.04$ | $0.081 \pm 0.03$ | $0.035 \pm 0.02$ |
| $P(k)^a$ | $0.77 \pm 0.13$ | $0.30 \pm 0.04$ | $0.12 \pm 0.03$ | $0.068 \pm 0.012$ | $0.024 \pm 0.010$ |
| Bouchet$^b$ et al. | $0.79 \pm 0.07$ | $0.26 \pm 0.08$ | $0.14 \pm 0.09$ | $0.087 \pm 0.100$ | $0.046 \pm 0.100$ |
| Mean: | $0.78 \pm 0.05$ | $0.29 \pm 0.03$ | $0.13 \pm 0.02$ | $0.074 \pm 0.01$ | $0.033 \pm 0.009$ |
| QDOT: | | | | | |
| Moore et al. (1 $\sigma$ errors) | $0.82^{+0.20}_{-0.20}$ | $0.34^{+0.11}_{-0.09}$ | $0.18^{+0.06}_{-0.07}$ | $0.12^{+0.05}_{-0.06}$ | $0.05^{+0.03}_{-0.03}$ |
| Efstathiou et al. (95% C.L.) | $0.87^{+0.23}_{-0.19}$ | $0.42^{+0.15}_{-0.11}$ | $0.26^{+0.12}_{-0.09}$ | $0.21^{+0.11}_{-0.07}$ | $0.047^{+0.063}_{-0.03}$ |
| Saunders$^c$ et al. (1 $\sigma$ errors) | | | $\sigma^2(14.5) = 0.44 \pm 0.091$ $\sigma^2(29.0) = 0.18 \pm 0.061$ $\sigma^2(58.0) = 0.067 \pm 0.019$ | | |

[a] cf., Fisher *et al.* (1993a)
[b] Converted from spherical cells.
[c] Converted from Gaussian density field.

# A  APPENDIX: MAXIMUM LIKELIHOOD FITTING TO $\xi(r)$

The two point correlation function is often characterized in the literature as a power law, $\xi(r) = (r_o/r)^\gamma$. Most authors fit power law models to their measured correlation functions by choosing the model parameters $(r_o, \gamma)$ which minimize the value of $\chi^2$ defined as

$$\chi^2 = \sum_{i=1}^{N} \frac{(\xi(r_i) - (r_o/r_i)^\gamma)^2}{\sigma_i^2} \quad , \tag{22}$$

where $\xi(r_i)$ and $\sigma_i^2$ are the measured values of the correlation function and its variance (e.g., from bootstrap resampling) at a separation $r_i$.

Strictly speaking, the process of $\chi^2$ minimization is valid only if the errors of the correlation function at a fixed separation are Gaussian distributed and if the values at different separations are *uncorrelated*. The power law model is usually fit to the data for $r \lesssim 20 \ h^{-1}$Mpc; since these separations are typically much smaller than the size of the sample, the central limit theorem ensures that the distribution of errors will be approximately Gaussian. In order to check this, we computed the correlation function for 100 mock *IRAS* observers for the CDM simulations discussed in § 2.2. The skewness in the distribution of the 100 correlation functions was consistent (within the expected $1\sigma$ errors) with zero for scales $s \gtrsim 2 \ h^{-1}$Mpc, although on smaller scales the distributions showed significant positive skewness. Thus, the assumption of Gaussian errors in the correlation function is reasonable.

However, because of the correlation between values of $\xi(r)$ at different separations, the probability that a set of measured values will be consistent with a given power-law model will not be described by the product of the individual probabilities of measuring each point, but rather by a $N_D$ dimensional multivariate Gaussian, where $N_D$ is the number of data points.

Here, we use principal component analysis (e.g., Kendall 1975) to derive a set of linear combinations of the measured values which are statistically independent, by finding the matrix which diagonalizes the covariance matrix of the measured values. Let us assume that we have measured the correlation function at $N_D$ different separations and that at *each* separation, $r$, we have $N$ different estimates of $\xi(r)$ obtained by bootstrap realizations of the correlation function; these estimates will be denoted by $\xi^{(j)}(r_i)$ for the $j^{\rm th}$ estimate at a separation $r_i$. For convenience, these variables can be made to have zero mean and unit variance,

$$x_i^{(j)} \equiv \frac{\xi^{(j)}(r_i) - \frac{1}{N}\sum_{k=1}^{N} \xi^{(k)}(r_i)}{\left[\frac{1}{N}\sum_{k=1}^{N}(\xi^{(k)}(r_i))^2 - \left(\frac{1}{N}\sum_{k=1}^{N} \xi^{(k)}(r_i)\right)^2\right]^{\frac{1}{2}}} \quad . \tag{23}$$

The covariance matrix of the $x_i^{(j)}$ variables is given by

$$\text{Cov}(i,j) = \frac{1}{N}\sum_{k=1}^{N} x_i^{(k)} x_j^{(k)} \quad . \tag{24}$$

Because the points are correlated, the covariance matrix has nonvanishing off-diagonal elements. However, if the covariance matrix has non-vanishing determinant, its symmetry guarantees the existence of a diagonalizing matrix, $R$. The columns of matrix $R$ are composed of the eigenvectors of the covariance matrix. Thus the covariance matrix is diagonalized by a unitary transformation. Given the matrix $R$, one can define the linear combinations $\tilde{x}_i^{(m)} \equiv R_{ik}^T x_k^{(m)}$, which are linearly independent, and thus can be used in a simple $\chi^2$ statistic, which we call $\mathcal{L}$.

If the residuals in the correlation function and the power law model with parameters $r_o$ and $\gamma$ have Gaussian distributions, then $\mathcal{L}$ will be distributed like $\chi^2$ with $\nu = N_D - 2$ degrees of freedom about its minimum value. The absolute goodness of fit of the model is given by the probability, $Q(\chi^2|\nu)$, that a $\chi^2$ distribution with $\nu$ degrees of freedom will exceed the observed value $\chi^2$ by chance. $Q(\chi^2|\nu)$ is given by the incomplete gamma function (cf., Press *et al.* 1992),

$$Q(\chi^2|\nu) = \frac{1}{\Gamma(\nu/2)} \int_{\chi^2/2}^{\infty} e^{-t} t^{\nu/2-1} dt \quad . \tag{25}$$

Values of $Q$ near unity indicate the model is an adequate representation of the data.

The joint confidence intervals for $r_o$ and $\gamma$ can be determined by computing $\mathcal{L}$ for a grid of values of $r_o$ and $\gamma$. The 68, 90, and 99% confidence intervals for the model parameters measured jointly are the areas enclosed by the contours $\Delta\mathcal{L} = 1.15, 2.305,$ and $4.605$, corresponding to the $\Delta\chi^2$ confidence intervals for a model with two parameters (cf., Press *et al.* 1992). The confidence interval for each parameter can be derived by projecting the confidence intervals appropriate for one degree of freedom (e.g., $\Delta\chi^2 = 1$ for the 68% level) onto the axis that corresponds to the desired parameter.